\documentclass[twocolumn,showpacs,preprintnumbers,amsmath,amssymb,superscriptaddress, bibnotes]{revtex4}

\usepackage{graphicx}
\usepackage{dcolumn}
\usepackage{bm}

\begin{document}
\draft
\preprint{HEP/123-qed}

\title{Itinerant Magnetism in URhGe Revealed by Angle Resolved Photoelectron Spectroscopy}

\author{Shin-ichi~Fujimori}
\affiliation{Condensed Matter Science Division, Japan Atomic Energy Agency, Sayo, Hyogo 679-5148, Japan}

\author{Ikuto~Kawasaki}
\altaffiliation[Present address: ]{Advanced Meson Science Laboratory, RIKEN Nishina Center for Accelerator Based Science, RIKEN, Wako, Saitama 351-0198, Japan}
\affiliation{Condensed Matter Science Division, Japan Atomic Energy Agency, Sayo, Hyogo 679-5148, Japan}

\author{Akira~Yasui}
\altaffiliation[Present address: ]{Japan Synchrotron Radiation Research Institute/SPring-8, Sayo, Hyogo 679-5198, Japan}
\affiliation{Condensed Matter Science Division, Japan Atomic Energy Agency, Sayo, Hyogo 679-5148, Japan}

\author{Yukiharu~Takeda}
\affiliation{Condensed Matter Science Division, Japan Atomic Energy Agency, Sayo, Hyogo 679-5148, Japan}

\author{Tetsuo~Okane}
\affiliation{Condensed Matter Science Division, Japan Atomic Energy Agency, Sayo, Hyogo 679-5148, Japan}

\author{Yuji~Saitoh}
\affiliation{Condensed Matter Science Division, Japan Atomic Energy Agency, Sayo, Hyogo 679-5148, Japan}

\author{Atsushi~Fujimori}
\affiliation{Condensed Matter Science Division, Japan Atomic Energy Agency, Sayo, Hyogo 679-5148, Japan}
\affiliation{Department of Physics, University of Tokyo, Hongo, Tokyo 113-0033, Japan}

\author{Hiroshi~Yamagami}
\affiliation{Condensed Matter Science Division, Japan Atomic Energy Agency, Sayo, Hyogo 679-5148, Japan}
\affiliation{Department of Physics, Faculty of Science, Kyoto Sangyo University, Kyoto 603-8555, Japan}

\author{Yoshinori~Haga}
\affiliation{Advanced Science Research Center, Japan Atomic Energy Agency, Tokai, Ibaraki 319-1195, Japan}

\author{Etsuji~Yamamoto}
\affiliation{Advanced Science Research Center, Japan Atomic Energy Agency, Tokai, Ibaraki 319-1195, Japan}

\author{Yoshichika~\=Onuki}
\affiliation{Advanced Science Research Center, Japan Atomic Energy Agency, Tokai, Ibaraki 319-1195, Japan}
\affiliation{Faculty of Science, University of the Ryukyus, Nishihara, Okinawa 903-0213, Japan}

\date{\today}

\begin{abstract}
The electronic structure of the ferromagnetic superconductor URhGe in the paramagnetic phase has been studied by angle-resolved photoelectron spectroscopy using soft x rays ($h\nu$=595-700~eV).
Dispersive bands with large contributions from U~5$f$ states were observed in the ARPES spectra, and form Fermi surfaces.
The band structure in the paramagnetic phase is partly explained by the band-structure calculation treating all U~5$f$ electrons as being itinerant, suggesting that an itinerant description of U~5$f$ states is a good starting point for this compound.
On the other hand, there are qualitative disagreements especially in the band structure near the Fermi level ($E_{\mathrm{B}} \lesssim 0.5$~eV).
The experimentally observed bands are less dispersive than the calculation, and the shape of the Fermi surface is different from the calculation. 
The changes in spectral functions due to the ferromagnetic transition were observed in bands near the Fermi level, suggesting that the ferromagnetism in this compound has an itinerant origin.

\end{abstract}

\pacs{79.60.-i, 71.27.+a, 71.18.+y}
\maketitle
\narrowtext
\section{INTRODUCTION}
The coexistence of magnetism and superconductivity in heavy fermion compounds is one of the central issues in condensed matter physics\cite{HF_Mag_SC}.
In particular, URhGe and UCoGe have attracted much attention in recent years because they show the coexistence of long-range ferromagnetic order and superconductivity under an ambient pressure\cite{AokiReview}.
They are weak ferromagnets, with $T_{\mathrm{Curie}}=9.5$~K and $M_0=0.4$~$\mu_{\mathrm{B}}$ (URhGe), and $T_{\mathrm{Curie}} \sim 3$~K and $M_0=0.05$~$\mu_{\mathrm{B}}$ (UCoGe).
They undergo transition into superconducting states below $T_{\mathrm{SC}}=0.26$~K (URhGe) \cite{URhGe1} and $T_{\mathrm{SC}}=0.7$~K (UCoGe) \cite{UCoGe}, and the superconducting state and magnetic orderings coexist below $T_{\mathrm{SC}}$.
The natures of the superconductivity and magnetism in these compounds have been well studied experimentally.
For example, the magnetic field and the pressure phase diagrams of these compounds have been obtained experimentally, and it has been suggested that these compounds are located near the quantum critical point of the magnetic transition\cite{URhGe2, URhGe3, UCoGe_phase}.
Furthermore, the NQR study of UCoGe clarified that the superconducting state and long-range magnetic ordering uniformly coexist in this compound \cite{UCoGe_NQR}.

On the other hand, the electronic structures of these compounds are not well understood.
An itinerant nature of U~5$f$ states has been expected from the small ordering moments in the ferromagnetic phase as well as their pressure dependences \cite{URhGe_LSDA1}.
In addition, x-ray photoelectron spectroscopy studies of URhGe \cite{Ucore}, URh$_{1-x}$Ru$_x$Ge \cite{URhRuGe_XPS}, and UCoGe \cite{UCoGe_XPS, Ucore} showed that the U~5$f$ state forms peak structures near the Fermi level ($E_{\mathrm{F}}$), suggesting that the U~5$f$ electrons in these compounds have an itinerant nature.
However, precise information about U~5$f$ states has not been obtained in these studies, and a definitive conclusion has not yet been obtained.
To establish realistic models of superconductivity and magnetism in these compounds, it is essential to reveal their microscopic electronic structures.

Here we report the direct observation of the band structure of URhGe through an angle-resolved photoelectron spectroscopy (ARPES) study. 
The band structure of URhGe is obtained for the first time, and it is found that the U~5$f$ electrons form Fermi surfaces (FS’s) in this compound.

\section{EXPERIMENTAL PROCEDURES}
Photoemission experiments were performed at the soft x-ray beamline BL23SU of SPring-8 \cite{BL23SU,BL23SU2}.
The overall energy resolution in angle-integrated photoemission (AIPES) experiments at $h\nu $=800~eV was about 110~meV, and that in ARPES experiments at $h\nu = 595--700$~eV was 100--160~meV, depending on the photon energies.
The position of $E_{\mathrm{F}}$ was carefully determined by measurements of the evaporated gold film.
Clean sample surfaces were obtained by cleaving the sample {\it in situ} with the surface parallel to the $ab$ plane.
The position of ARPES cuts was calculated by assuming a free-electron final state with an inner potential of $V_{0}$=12~eV.

\section{RESULTS AND DISCUSSION}
\subsection{Angle-integrated photoemission spectra}
\begin{figure}
\includegraphics[scale=0.45]{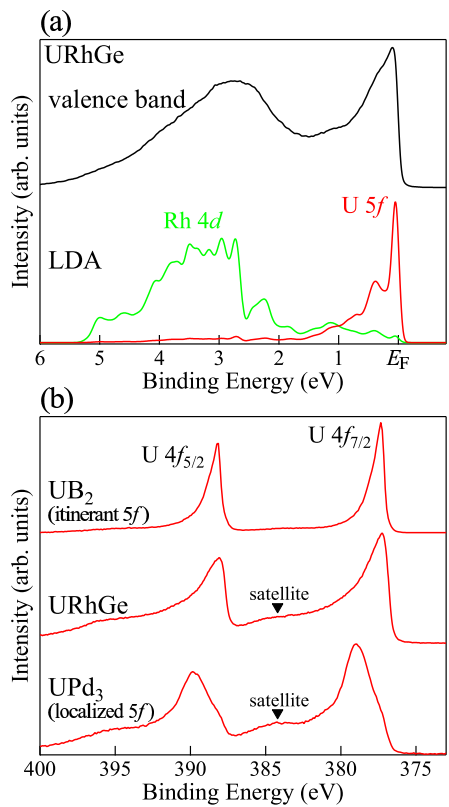}
\caption{(Online color)
Angle-integrated photoemission spectra of URhGe measured with $h\nu $=800~eV.
(a) Valence-band spectra of URhGe and calculated partial density of states of the Rh~4$d$ and U~5$f$ states.
(b) U~4$f$ core-level spectrum of URhGe together with those of the typical itinerant compound UB$_2$ and localized compound UPd$_3$.}
\label{AIPES}
\end{figure}
First, we present the AIPES spectra of URhGe.
Figure~\ref{AIPES} (a) shows the valence-band spectrum of URhGe taken at $h\nu $=800~eV.
The sample temperature was kept at 20~K, and the compound is in a paramagnetic phase.
The spectrum is identical to that in Ref.~\cite{Ucore}.
In this photon energy range, the contributions from U~5$f$ and Rh~4$d$ states are dominant, and those from $s$ and $p$ states are two or three orders of magnitude smaller than those of U~5$f$ and Rh~4$d$ states \cite{Atomic}.
In the valence-band spectrum, there is a sharp peak structure just below $E_{\mathrm{F}}$.
On the high-binding-energy side, there is a broad peak structure distributed at $2--5$~eV.
To understand the origin of this peak structure, we have compared this spectrum with the result of the band-structure calculation.
In the calculation, relativistic-linear-augmented-plane-wave (RLAPW) band-structure calculations\cite{Yamagami} within the local density approximation (LDA)\cite{LDA} were performed for URhGe treating all U~5$f$ electrons as being itinerant.
In the lower part of the Fig.~\cite{AIPES} (a), the calculated U~5$f$ and Rh~4$d$ density of states broadened by the instrumental resolution are indicated.
Comparison between the spectrum and the calculated density of states suggests that the sharp peak structure near $E_{\mathrm{F}}$ and the broad peak structure on the high-binding-energy sides correspond to U~5$f$ and Rh~4$d$ states, respectively. 
Here, it should be noted that a shoulder-like structure is recognized at around $E_{\mathrm{B}} =0.5$~eV in the experimental spectrum.
A similar structure is present in the calculated U~5$f$ DOS as well, and this suggests that this shoulder structure originates not from the electron correlation effect, but from the band structure. 
The overall spectral shape is consistent with the results of the band-structure calculation.

Figure~\ref{AIPES} (b) shows the U~4$f$ core-level spectrum of URhGe together with those of the typical itinerant 5$f$ compound UB$_2$ and localized 5$f$ compound UPd$_3$.
These data are taken from Ref.\cite{Ucore}.
They show a spin-orbit splitting corresponding to U~4$f_{7/2}$ and U~4$f_{5/2}$, and both of them have asymmetric line shapes.
The core-level spectrum of URhGe has a relatively simple spectral line shape.
It is similar to that of the itinerant 5$f$ compound UB$_2$, suggesting that U~5$f$ electrons in URhGe have an itinerant character. 
Meanwhile, the spectrum of URhGe is much broader than that of UB$_2$, and is accompanied by a satellite structure on the high-binding-energy side, as has been observed in the UPd$_3$ spectrum.
This implies that the U~5$f$ electrons in URhGe have a correlated character as well.

\subsection{Band structure in the paramagnetic phase}
\begin{figure}
\includegraphics[scale=0.35]{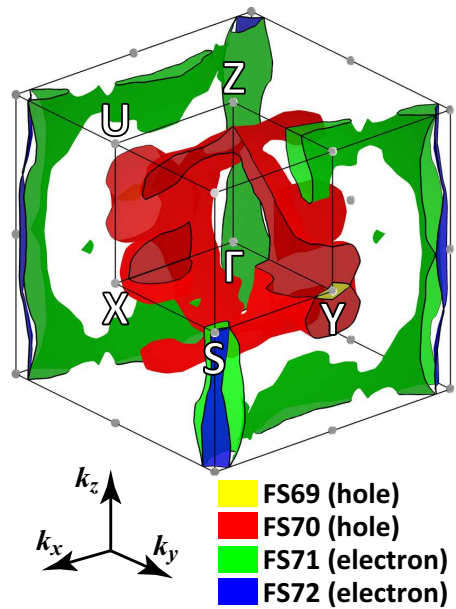}
\caption{(Online color)
Brillouin zone of URhGe and calculated Fermi surfaces.}
\label{3DFS}
\end{figure}
We first explain about the Brillouin zone and the calculated FSs of URhGe.
Figure~\ref{3DFS} shows the simple orthorhombic Brillouin zone of URhGe and the calculated FSs in the paramagnetic phase.
In the band-structure calculation, bands 69--72 form FSs as shown in the figure.
FS~69 is a small hole pocket around the Y point.
FS~70 has a highly three-dimensional shape.
It forms a connected hole FS along the $k_y$ direction and a hole pocket FS in the middle of the $\mathrm{\Gamma}$ and X points.
FS~71 is an electron FS with a grid-like shape spreading along the $k_x$--$k_z$ plane.
FS~72 has a pillar-like shape along the $k_z$ direction at the corner of the Brillouin zone.
All these calculated FSs have large contributions from U~5$f$ states.
The calculation suggests that the electronic structure of this compound is highly three-dimensional in nature.

\begin{figure*}
\includegraphics[scale=0.4]{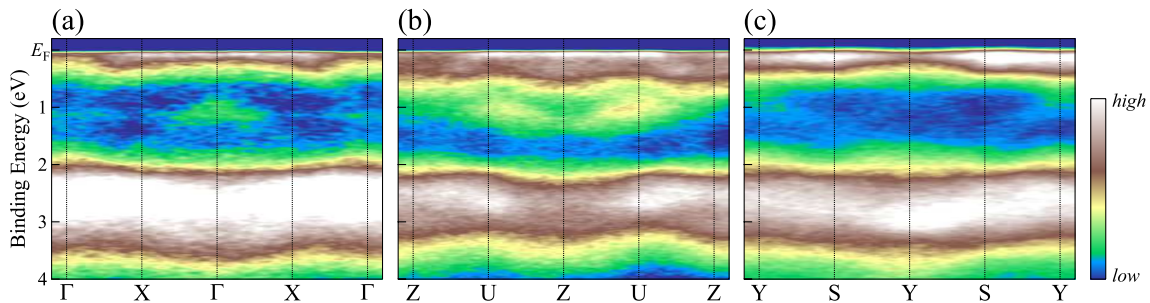}
\caption{(Online color)
ARPES spectra of URhGe in the paramagnetic phase along selected high-symmetry lines.
(a) ARPES intensity map along the X-$\mathrm{\Gamma}$-X line ($h\nu$=625~eV).
(b) ARPES intensity map along the U-Z-U line ($h\nu$=625~eV).
(c) ARPES intensity map along the S-Y-S line ($h\nu$=700~eV).}
\label{ARPES1}
\end{figure*}
Figure~\ref{ARPES1} shows the ARPES spectra of URhGe measured along the X-$\mathrm{\Gamma}$-X [Fig.~\ref{ARPES1}(a)], U-Z-U [Fig.~\ref{ARPES1}(b)], and S-Y-S [Fig.~\ref{ARPES1}(c)] high-symmetry lines.
The sample temperature was kept at 20~K, which was in the paramagnetic phase.
The position of the ARPES cut in the momentum space is calculated based on the free electron final sates, and the photon energies used were $h \nu$=625~eV for the X-$\mathrm{\Gamma}$-X and U-Z-U lines, and $h \nu$=700~eV for the S-Y-S line.
In these ARPES spectra, clear energy dispersions were observed.
In the vicinity of $E_{\mathrm{F}}$, there exist dispersive bands with strong intensities.
These are contributions from the U~5$f$ quasiparticle bands.
The peak intensities of these structures have strong momentum dependences, suggesting that the U~5$f$ quasiparticle bands have finite energy dispersions.
On the high-binding-energy side ($E_{\mathrm{B}}=2-4$~eV), there exist dispersive bands with strong intensities.
These are the contributions from the Rh~4$d$ bands.
\begin{figure*}
\includegraphics[scale=0.4]{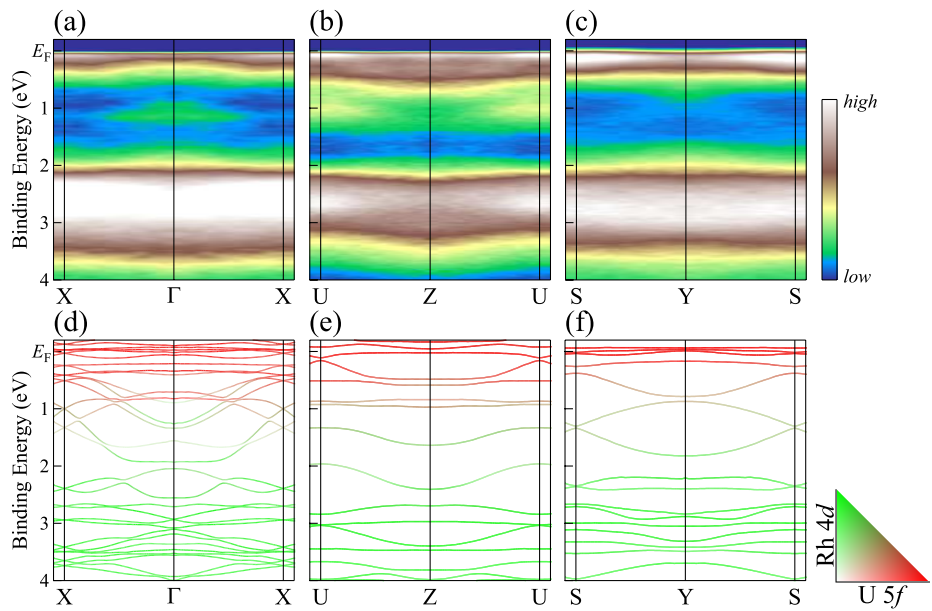}
\caption{(Online color)
Symmetrized ARPES spectra and results of band-structure calculation.
(a-c) Symmetrized ARPES spectra.
(d-f) Results of band-structure calculation.
The color coding of bands is the projection of the contributions from U~5$f$ states and Rh~4$d$ states respectively.}
\label{ARPES2}
\end{figure*}

Here, it should be noted that those bands are not symmetric relative to the high-symmetry points.
For example, the spectra measured along the X-$\mathrm{\Gamma}$-X high-symmetry line are not symmetric relative to the X point in Fig.~\ref{ARPES1} (a).
This may be due to the photoemission structure factor effect as has been observed in ARPES spectra of other materials \cite{PSF, UN_ARPES}.
To eliminate this effect, we have symmetrized these ARPES spectra relative to the high-symmetry points.
Figures ~\ref{ARPES2} (a)-\ref{ARPES2}(c) show the symmetrized ARPES intensity maps.
The behaviors of the bands are clearly demonstrated in these images.

To evaluate the validity of the itinerant description of the U~5$f$ states in this compound, we compare the present ARPES spectra with the result of the band-structure calculation treating all U~5$f$ electrons as being itinerant.
Figures~\ref{ARPES2} (d)-\ref{ARPES2} (f) show the calculated band structure.
The color coding is the projection of the contributions from U~5$f$ states and Rh~4$d$ states respectively.
The contributions from U~5$f$ states are distributed in the energy range of $E_{\mathrm{B}} < 1$~eV, while those from Rh~4$d$ states are mainly distributed in the energy range of $E_{\mathrm{B}} > 2$~eV.
Many dispersive bands exist in the calculation, and it is difficult to compare them with the experimentally observed bands one by one.
Meanwhile, the overall band structures have some similarities between the experiment and the calculation.
On the high-binding-energy side ( $E_{\mathrm{B}} >2$~eV), the contributions from Rh~4$d$ states were observed in both the experiment and the calculation.
In the vicinity of $E_{\mathrm{F}}$, there are weakly dispersive bands in both the experiment and the calculation, and these are contributions from U~5$f$ states.
There exist few bands in the energy region in between them ($E_{\mathrm{B}} =0.8-2$~eV), and some calculated bands seem to correspond to the experimental spectra.
For example, there is an inverted parabolic band centered at the Y point in the energy region of $E_{\mathrm{B}} = 0.7-1.3$~eV in the calculation [Fig.~\ref{ARPES2} (f) ] that exists in the experiment [Fig.~\ref{ARPES2} (c) ].
The similar inverted parabolic bands centered at the $\mathrm{\Gamma}$ point in the energy region of $E_{\mathrm{B}} = 0.9-1.3$~eV (Fig.~\ref{ARPES2} (d) ) are seen in the experimental spectra [ Fig.~\ref{ARPES2}(a) ].

\begin{figure*}
\includegraphics[scale=0.4]{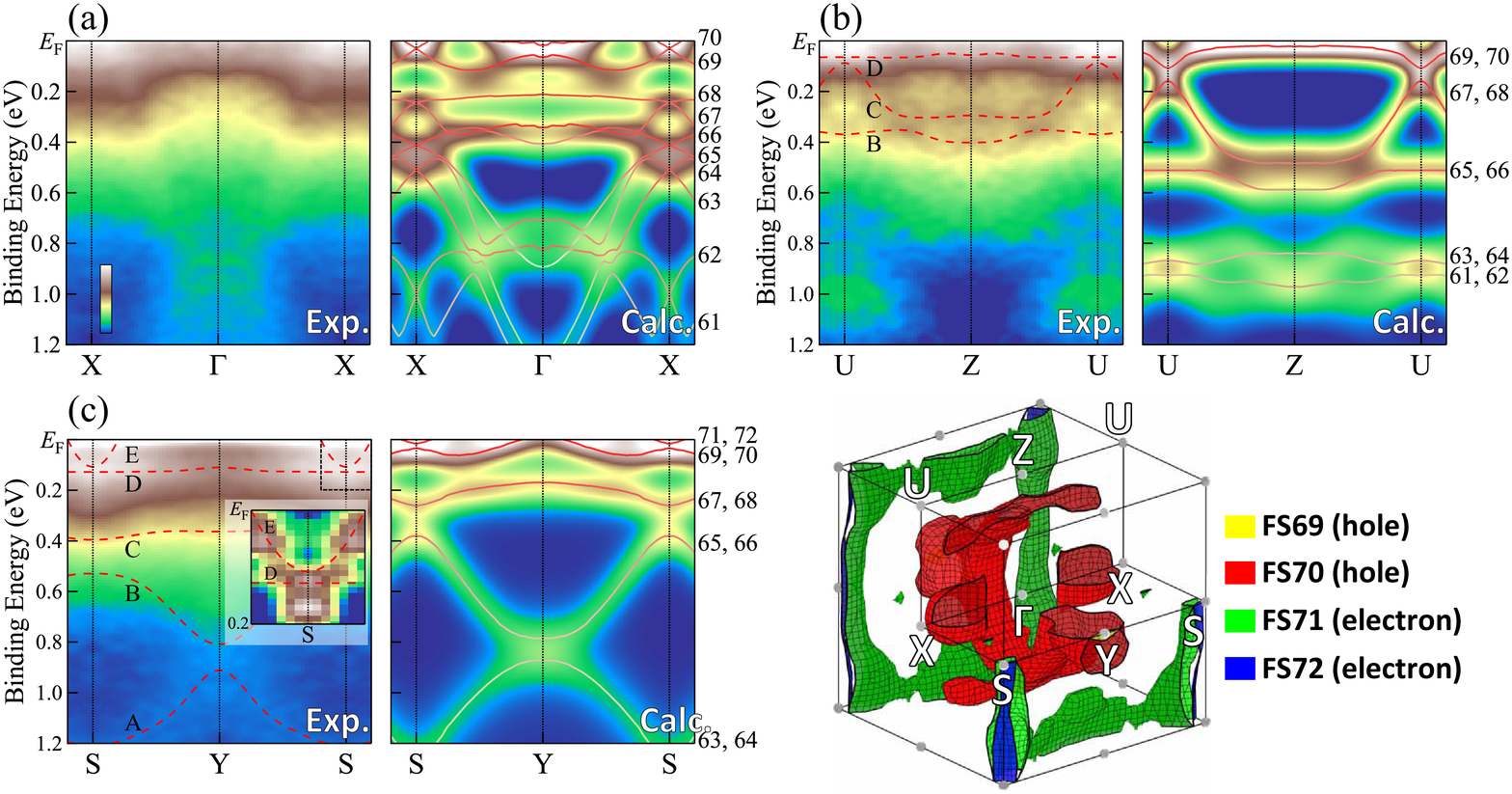}
\caption{(Online color)
Blowup of experimental ARPES spectra and their simulation based on band-structure calculations along the (a) X-$\mathrm{\Gamma}$-X, (b) U-Z-U, and (c) S-Y-S high-symmetry lines.
The color coding of bare calculated bands shown by the solid lines is the same as in Figs.~\ref{ARPES2} (d)-\ref{ARPES2} (f).
These spectra are divided by the Fermi-Dirac function broadened by the experimental energy resolution.
Inset in (c): ARPES spectra along the S-Y-S high-symmetry line shows the spectra normalized by the area of the momentum distribution curves.
}
\label{ARPES3}
\end{figure*}
To see details of the band structure near $E_{\mathrm{F}}$ as well as its correspondence to the band-structure calculation, a blowup of the experimental ARPES spectra and their simulation based on band-structure calculation are shown in Fig.~\ref{ARPES3}.
Figure~\ref{ARPES3} shows a comparison between experimental ARPES spectra and their simulations along the X-$\mathrm{\Gamma}$-X [ Fig.~\ref{ARPES3} (a) ], U-Z-U [ Fig.~\ref{ARPES3} (b) ], and S-Y-S high-symmetry lines [ Fig.~\ref{ARPES3} (b) ].
These spectra are divided by the Fermi-Dirac function broadened by the instrumental energy resolution to observe the states near $E_{\mathrm{F}}$ more clearly.
The approximate positions of experimental bands are estimated by the second derivatives of the energy distribution curves or momentum distribution curves, and are shown by dashed lines along the U-Z-U high-symmetry line [Fig.~\ref{ARPES3} (b) ] and the S-Y-S high-symmetry lines [ Fig.~\ref{ARPES3} (c) ].
In the simulation, the following effects were taken into account: (i) the broadening along the $k_z$ direction due to the finite escape depth of photoelectrons, (ii) the lifetime broadening of the photo-hole, (iii) the photoemission cross sections of orbitals, and (iv) the energy resolution and the angular resolution of the electron analyzer.
The details are described in Ref.~\cite{UN_ARPES}.

The correspondence between the ARPES spectra and the calculations is more clearly recognized.
A detailed comparison suggests that there are some similarities between the experiment and the calculation.
Some agreements are clearly identified, especially in the spectra along the U-Z-U and S-Y-S high-symmetry lines.
For example, along the U-Z-U high-symmetry line, the experimentally observed three bands B, C, and D, correspond to the calculated bands 65--66, 67--68, and 69--70, respectively.
In addition, there are similar qualitative agreements in the spectra along the S-Y-S high-symmetry line.
Bands A, B, C, D, and E in the experimental spectra correspond to the bands 63--64, 65--66, 67--68, 69-70, and 71--72, respectively.
In particular, band E forms a small electron pocket around the S point, which can be clearly recognized in the spectra normalized by the areas of the momentum distribution curves shown in the inset in the Fig.~\ref{ARPES3} (c).
A similar electron pocket exists in the band-structure calculation as bands 71 and 72.
On the other hand, band D does not have a large energy dispersion, and it does not form a FS as bands 69 and 70 do in the calculation.
Therefore, the agreement is unsatisfactory, especially in bands near $E_{\mathrm{F}}$.
The agreement is further unclear in the spectra along the X-$\mathrm{\Gamma}$-X high-symmetry line. 
The calculated spectra have a complicated structure, and it seems very different from the experimental spectra.
However, there are still some corresponding features in both the experiment and the calculation.
For example, the inverted parabolic band with its apex at $E_{\mathrm{B}} \sim$~0.2~eV at the X point in the experimental spectra has a correspondence to bands 63--64 in the calculation. 
The inverted parabolic band with its apex at $E_{\mathrm{B}} \sim$~1.0~eV at the $\mathrm{\Gamma}$ point corresponds to part of the calculated bands 61--65. 
The states near $E_{\mathrm{F}}$ in the experimental spectra are rather featureless, and they are very different from the calculation.
Therefore, the agreement between the experiment and the calculation is limited, especially in states near $E_{\mathrm{F}}$, and we could not obtain information on the shape of FS's from the present experimental data.
In addition, many flat bands are expected around $E_{\mathrm{F}}$ in the calculation, and even a tiny change in $E_{\mathrm{F}}$, by the order of 10~meV, drastically changes the shape of calculated FS’s. 
This makes it more difficult to compare states near $E_{\mathrm{F}}$ between the experiment and the calculation.

\subsection{Band structure in the ferromagnetic phase}
\begin{figure*}
\includegraphics[scale=0.4]{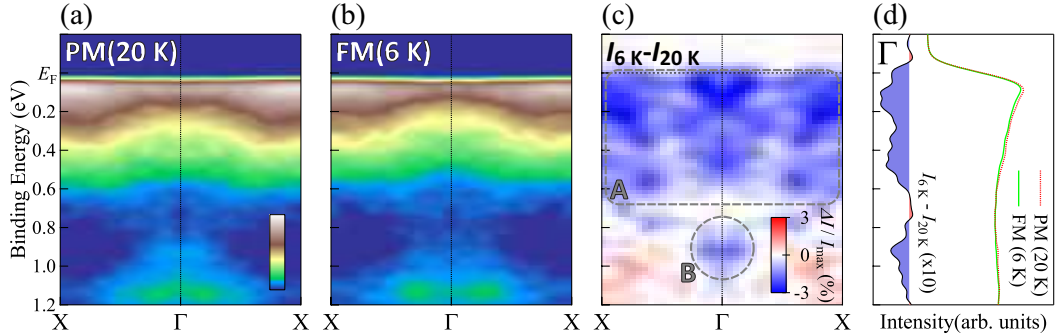}
\caption{(Online color)
Temperature dependence of ARPES spectra of URhGe.
(a) ARPES intensity map along the X-$\mathrm{\Gamma}$-X direction measured at 20~K (paramagnetic phase).
(b) ARPES spectra measured at 6~K (ferromagnetic phase).
(c) Image plot of the difference in ARPES spectra measured at 20~K vs 6~K.
Intensities are normalized with the highest intensity in the momentum and energy area shown in this figures.
(d) Comparison of ARPES spectra measured at the $\mathrm{\Gamma}$ point at 20~K vs 6~K and their difference.
}
\label{tempdep}
\end{figure*}
Next, we show the changes in the electronic structure due to the ferromagnetic transition.
Figures~\ref{tempdep} (a) and \ref{tempdep} (b) show blowups of ARPES spectra measured along the X-$\mathrm{\Gamma}$-X line at 20~K (paramagnetic phase) and 6~K (ferromagnetic phase), respectively.
Although the changes are not significant, some temperature dependencies were clearly observed.
Here, it should be noted that the energy difference between these two sample temperatures (20~K, $\sim$1.7~meV; 6~K, $\sim$0.5~meV ) is much smaller than the energy scale of the changes in the ARPES spectra, suggesting that the changes are not due to a thermal broadening effect.
To see details of the changes, we have subtracted the spectra measured at 20~K from those measured at 6~K.
The intensity map of the difference in the spectra measured at 20~K vs 6~K is shown in Fig.~\ref{tempdep} (c), and the spectra at the $\mathrm{\Gamma}$ point are depicted in Fig.~\ref{tempdep} (d).
Intensities are normalized with the highest intensity in the momentum and energy area shown in this figure.
Both spectra are normalized to the intensities of the Rh 4$d$ bands located at $E_{\mathrm{B}} >1.5$~eV.
There are two kinds of changes in the spectral functions.
First, the intensities in the region of $E_{\mathrm{B}} <0.6$~eV, which correspond to contributions mainly from U~5$f$ quasiparticle bands, decrease in the ferromagnetic phase.
The area is designated A in the figure.
The changes in spectral intensities near $E_{\mathrm{F}}$ suggest that the ferromagnetic ordering in this compound originates from the itinerant U~5$f$ quasiparticle bands.
In addition to these changes, the intensities at around $E_{\mathrm{B}} \sim 1.0$~eV at the $\mathrm{\Gamma}$ point decrease.
The area is designated B in the figure.
The origin of this change is discussed in the next section.

\subsection{Discussion}
Accordingly, we have observed dispersive U~5$f$ quasiparticle bands near $E_{\mathrm{F}}$ in ARPES spectra of URhGe.
They form FSs in this compound, suggesting that they have basically an itinerant nature.
On the other hand, the agreement between the experimentally obtained spectra and the band-structure calculation is limited.
The agreement is not satisfactory as those of the itinerant paramagnets UFeGa$_5$ \cite{UFeGa5_ARPES} and UB$_2$ \cite{UB2_ARPES}, or the itinerant antiferromagnet UN \cite{UN_ARPES}.
In particular, the states near $E_{\mathrm{F}}$ ($E_{\mathrm{B}} \lesssim 0.5$~eV) show considerable deviations from the calculation.
The ellipsoidal pocket FS with a size of about 7 \% of the Brillouin zone was observed in SdH oscillations \cite{URhGe3}, but it does not exist in the calculated FS’s.
This also suggests that the shape of FS’s might be different from the calculation.
These claim that the LDA might be a reasonable starting point to describe its electronic structure, but inclusion of the electron correlation is needed to describe its electronic structure. 
Furthermore, the core-level spectrum of URhGe has a satellite peak on the high-binding-energy side of the main line as shown in Fig.~\ref{AIPES} (b), suggesting the importance of the dynamical screening effect in URhGe.
It has been shown that the inclusion of dynamical correlation effects alters the band structure near $E_{\mathrm{F}}$ \cite{dmft}, and such a theoretical framework is a step forward to the understanding of this compound.

Changes in the spectral line shape due to the ferromagnetic transition were observed in the vicinity of $E_{\mathrm{F}}$ as well as on the high-binding-energy side ($E_{\mathrm{B}} \sim 1$~eV).
The former change suggests that the ferromagnetic ordering in this compound is due to the changes in itinerant quasiparticle bands near $E_{\mathrm{F}}$.
Although their details were not resolved in the present spectra, the changes are presumably due to the splitting of bands into majority-spin and minority-spin bands in the ferromagnetic phase as has been observed in UTe \cite{UTe} and UIr \cite{UIr} since U~5$f$ electrons have an itinerant nature in URhGe as well.
In the Stoner-type mean-field model of itinerant ferromagnetism, the exchange splitting energy $\Delta_{\mathrm{ex}}$ is expressed as $\Delta_{\mathrm{ex}}=IM$ where $I$ and $M$ represent the Stoner parameter and the magnetization respectively.
The Stoner parameter $I$ might be somewhat different in the cases of UTe and URhGe, but the smaller magnetic moment of URhGe ($M_0=0.4$~$\mu_{\mathrm{B}}$) compared to UTe ($M_0=2.25$~$\mu_{\mathrm{B}}$) implies that the energy shift of the majority and minority bands ($\sim \Delta_{\mathrm{ex}}/2$) of URhGe should be much smaller than that of UTe ($\Delta_{\mathrm{ex}} /2 \sim$ 50~meV\cite{UTe}).
Furthermore, it is experimentally known that $\Delta_{\mathrm{ex}}$ approximately scales with $T_{\mathrm{Curie}}$ in itinerant ferromagnets\cite{CurieTmp}, and the one order-lower transition temperature of URhGe ($T_{\mathrm{Curie}}=9.5$~K) compared to UTe ($T_{\mathrm{Curie}}=104$~K) also argues that its $\Delta_{\mathrm{ex}}$ should be much smaller than that of UIr.
These imply that the $\Delta_{\mathrm{ex} / 2}$ of URhGe should be of the order of 10~meV, and its direct observation is difficult at the present energy resolution ($\Delta E \sim 100$~meV).

Meanwhile, the latter change on the high-binding-energy side is not the mean-field-like splitting of energy bands, but the change in peak structure.
This change might originate from the incoherent part of the spectrum due to the correlation effect of U~5$f$ states. 
Riseborough \cite{Riseborough} suggested that the incoherent spin excitation produces an incoherent peak in the off-$E_{\mathrm{F}}$ region in a weak ferromagnet near a quantum critical point.
A similar change was observed in the spectra of the ferromagnet UIr, where the peak at the $\Gamma$ point and $E_{\rm B} \sim$0.5~eV was changed, in addition to the state near $E_{\mathrm{F}}$ \cite{UIr}, by the ferromagnetic transition.
Therefore, this change on the high-binding-energy side might be a common feature of uranium ferromagnets located near a quantum critical point.

\section{CONCLUSION}
In conclusion, we have found that quasiparticle bands with large contributions from U~5$f$ states form FSs of URhGe.
The overall band structure of URhGe is explained by the band-structure calculation based on the LDA, but the shape of the band structure near $E_{\mathrm{F}}$ shows considerable deviations from the calculation.
The experimental band structure near $E_{\mathrm{F}}$ is rather featureless, and the shapes of FS's are qualitatively different from the calculation.
In addition, the U~4$f$ core-level spectrum of URhGe is accompanied by a satellite peak, which is a signature of the electron correlation effect.
These results suggest that the inclusion of an electron correlation effect is essential to describe its electronic structure although the band-structure calculation is an appropriate starting point.
The changes in ARPES spectra associated with the ferromagnetic transition were observed in the band near $E_{\mathrm{F}}$ as well as states on the high-binding-energy side.
The former should be due to the exchange splitting of quasiparticle bands, while the latter might be related to the correlated nature of U~5$f$ states.

\acknowledgments
We would like to thank P.~Riseborough for stimulating discussion and comments.
The experiment was performed under Proposal No. 2010B3824 at SPring-8 BL23SU.
The present work was financially supported by a Grant-in-Aid for Scientific Research from the Ministry of Education, Culture, Sports, Science, and Technology, Japan, under Contact No. 21740271; Grants-in-Aid for Scientific Research on Innovative Areas "Heavy Electrons" (Nos. 20102002 and 20102003) from the Ministry of Education, Culture, Sports, Science, and Technology, Japan; and the Shorei Kenkyu from Hyogo Science and Technology Association.


\end{document}